\documentclass[a4paper,12pt,titlepage]{article}
\usepackage{graphicx}
\begin{document}
\begin{titlepage}
\begin{flushright}
WU-HEP-99-2
\end{flushright}
\vspace{5mm}
\begin{center}
\Large\textbf{%
Stochastic Quantization of Bottomless Systems\\
---Stationary quantities in a diffusive process---}
\end{center}
\vspace{-3mm}
\begin{center}
\large%
Kazuya Yuasa\footnote{JSPS Research Fellow. Email address:
yuasa@hep.phys.waseda.ac.jp} and Hiromichi
Nakazato\footnote{Email address: nakazato@hep.phys.waseda.ac.jp}
\end{center}
\vspace{-5mm}
\begin{center}
\large\textit{%
Department of Physics, Waseda University, Tokyo 169-8555,
Japan\/%
}
\end{center}
\vspace{0mm}
\begin{center}
\ 
\end{center}
\vfil
\begin{center}
\small\textbf{Abstract}
\end{center}

{\small
By making use of the Langevin equation with a kernel, it was
shown that the Feynman measure $e^{-S}$ can be realized in a
restricted sense in a diffusive stochastic process, which
diverges and has no equilibrium, for bottomless systems.
In this paper, the dependence on the initial conditions and the
temporal behavior are analyzed for 0-dim bottomless systems.
Furthermore, it is shown that it is possible to find
\textit{stationary\/} quantities.}
\vfil
\end{titlepage}

\section{Introduction}
\setcounter{equation}{0}
It remains an unsettled question in quantum field theory how to
deal with bottomless systems, i.e.~systems whose Euclidean
actions are unbounded from below.
Important and familiar examples include Einstein
gravity \cite{ref:Gravity}.
The fact that one cannot define the vacuum for such systems makes
it impossible to quantize them in the usual ways.
The standard path integral quantization procedure, for example,
is not applicable, since the Feynman measure $e^{-S}$ with a
bottomless action $S$ is not normalizable and the vacuum
expectation values cannot be defined.

In the language of the stochastic quantization of Parisi and
Wu \cite{ref:SQM}, the difficulty with bottomless systems
manifests itself as the absence of thermal equilibrium, where
quantum theory is supposed to be realized for ordinary (normal)
systems.
In spite of this apparent drawback, however, it has been argued
by several people that the potentiality inherent in yet to be
studied stochastic dynamics may offer the possibility to properly
deal with bottomless systems.
Such attempts have already been started in
Refs.~\cite{ref:Greensite} and \cite{ref:Tanaka}, though on the
basis of quite different ideas.
An analytical expression of the probability distribution obtained
for 0-dim bottomless systems \cite{ref:Nakazato} supports the
idea expressed in Ref.~\cite{ref:Tanaka}, i.e.~realization of
the Feynman measure $e^{-S}$ in a finite space-time region at
finite fictitious time by means of the kernel-degree of freedom.
We must note that here the stochastic process is of the diffusion
type and has no equilibrium for bottomless systems, since the
desired distribution $e^{-S}$ is not normalizable and therefore
no longer belongs to the spectrum of the Fokker--Planck operator
$H$; every eigenstate of $H$ necessarily decays away in the
large-time limit.
Every quantity measured in this process is thus dependent on the
initial conditions and is dependent on a fictitious time, in
general.
This is a severe problem if one tries to extract from this kind
of treatment physically meaningful and sensible quantities, which
should, of course, be independent of any of the stochastic
dynamics.

The purpose of this paper is thus first to examine the dependence
of the probability distribution on the initial conditions and to
clarify the behavior of the correlation functions in this
framework.
It is observed that at large fictitious times the correlation
functions diffuse quite universally, irrespectively of the
initial distribution.
This suggests the possibility of extracting such a quantity that
does not depend on the initial conditions at large times.
We then propose a method that enables us to extract it from the
diffusive stochastic process.

\section{Stochastic quantization with bottomless systems}
\setcounter{equation}{0}
Let us first briefly review the stochastic quantization of Parisi
and Wu \cite{ref:SQM}.
In stochastically quantizing the system with a Euclidean action
$S[\phi]$, one sets up the Langevin equation
\begin{equation}
\frac{\partial}{\partial t}\phi(x,t)
=-\frac{\delta S[\phi]}{\delta\phi(x,t)}+\eta(x,t),
\label{eqn:Langevin}
\end{equation}
which governs the stochastic dynamics of $\phi$ with respect to
the fictitious time $t$.
Here $\eta$ is the Gaussian white noise characterized by the
statistical properties
\begin{equation}
\langle\eta(x,t)\rangle=0,\quad
\langle\eta(x,t)\eta(x',t')\rangle
=2\delta^D\!(x-x')\delta(t-t'),\quad
{\mathrm{etc.}}
\label{eqn:GaussianWhiteNoise}
\end{equation}
The quantum theory has been shown to be realized in the thermal
equilibrium limit $t\to\infty$.
One can equivalently work with the Fokker--Planck equation
\begin{equation}
\frac{\partial}{\partial t}P[\phi;t]=H[\phi]P[\phi;t],
\label{eqn:FokkerPlanck}
\end{equation}
with
\begin{equation}
H[\phi]=\int d^D\!x\,\frac{\delta}{\delta\phi(x)}
\left(\frac{\delta}{\delta\phi(x)}
+\frac{\delta S[\phi]}{\delta\phi(x)}\right),
\label{eqn:FokkerPlanckOperator}
\end{equation}
where $P[\phi;t]$ is the probability distribution of $\phi$ at
time $t$.
It is easy to show that the Fokker--Planck operator $H$ is
negative semi-definite for any $S$ and that the eigenfunctional
of $H$ belonging to the highest zero eigenvalue is $e^{-S}$,
which ensures the relaxation of $P$ to the distribution $\propto
e^{-S}$, irrespective of the choice of the initial distribution,
provided that the spectrum includes the discrete zero.
Quantum field theory is thus given in the thermal equilibrium
limit of a hypothetical stochastic process.

This is, however, not the case with bottomless systems, since
$e^{-S}$ is not normalizable and hence does not belong to the
spectrum of the Fokker--Planck operator $H$.
Eigenvalues are negative definite, and the hypothetical
stochastic process has no thermal equilibrium limit (i.e., it is
a diffusion process).

\section{Attempts to deal with bottomless systems}
\setcounter{equation}{0}
Although the naive application of stochastic quantization to
bottomless systems does not work, some attempts to deal with them
have been reported \cite{ref:Greensite,ref:Tanaka}.
Greensite and Halpern \cite{ref:Greensite} assumed that the
meaningful distribution function in a diffusive stochastic
process for bottomless systems is the highest normalizable
eigenstate of the Fokker--Planck operator.
That is, they gave up using the Feynman measure $e^{-S}$ to
evaluate expectation values, since $e^{-S}$, which is not
normalizable for the bottomless action $S$, cannot belong to the
spectrum of the Fokker--Planck operator.
Instead, they proposed to utilize its true ground state: the
Feynman measure $e^{-S}$ was abandoned.
Since the true ground state is not a stationary state and it
finally decays away at large times, they had to devise a way how
to extract it from the diffusive stochastic process.

On the other hand, Tanaka et al.~pursued the possibility of
producing the distribution $e^{-S}$ even for a bottomless action
$S$ by making use of the Langevin equation
\begin{equation}
\frac{\partial}{\partial t}\phi(x,t)
=-K[\phi]\frac{\delta S[\phi]}{\delta\phi(x,t)}
+\frac{\delta K[\phi]}{\delta\phi(x,t)}+K^{1/2}[\phi]\eta(x,t)
\label{eqn:KerneledLangevin}
\end{equation}
with a positive kernel functional $K$ \cite{ref:Tanaka}.
Here and in what follows, stochastic differential equations are
of the Ito-type \cite{ref:SDE}.
Note that the corresponding Fokker--Planck operator is given by
\begin{equation}
H[\phi]
=\int d^D\!x\,\frac{\delta}{\delta\phi(x)}K[\phi]
\left(
\frac{\delta}{\delta\phi(x)}+\frac{\delta S[\phi]}{\delta\phi(x)}
\right),
\label{eqn:KerneledFokkerPlanckOperator}
\end{equation}
which has negative semi-definite eigenvalues for any positive
kernel $K$, and that the thermal equilibrium distribution could
again be given by $e^{-S}$ \textit{only if it is normalizable\/}.
They expected that an appropriate choice of the positive kernel
$K$ may enable stochastic variables to be confined in a finite
region and that the desired distribution $e^{-S}$ could be
reproduced there.
Their numerical simulation of the Langevin equation
(\ref{eqn:KerneledLangevin}) with a specific choice of the kernel
$K$ for simple 0-dim models actually seems to support their
expectation.

In order to see the actual stochastic dynamics more clearly and
in detail, let us restrict ourselves to 0-dim cases.
The Langevin equation now reads
\begin{equation}
\dot{x}=-K(x)S'(x)+K'(x)+K^{1/2}(x)\eta,
\label{eqn:ZeroKerneledLangevin}
\end{equation}
and the corresponding Fokker--Planck equation is given by
\begin{equation}
\dot{P}(x,t)
=\frac{\partial}{\partial x}K(x)
\left(\frac{\partial}{\partial x}+S'(x)\right)P(x,t).
\label{eqn:ZeroKerneledFokkerPlanck}
\end{equation}
Here dots and primes denote differentiation with respect to the
fictitious time $t$ and $x$, respectively.
It is shown in Ref.~\cite{ref:Nakazato} that the Fokker--Planck
equation (\ref{eqn:ZeroKerneledFokkerPlanck}) can be solved
analytically with an appropriate choice of the kernel function
$K$ for any 0-dim action $S$ which is unbounded from below for
$x\to\pm\infty$.
The choice of the kernel $K$
\begin{equation}
K(x)=e^{2S(x)}
\label{eqn:Kernel}
\end{equation}
yields the Green function for the Fokker--Planck equation
(\ref{eqn:ZeroKerneledFokkerPlanck}) \cite{ref:Nakazato}
\begin{equation}
P(x,t;x_0)
=e^{-S(x)}\frac{1}{\sqrt{4\pi t}}
\exp\!\left(-\frac{f^2(x)}{4t}\right),\quad
f(x)=\int^x_{x_0}dy\,e^{-S(y)}.
\label{eqn:GreenFunction}
\end{equation}
This satisfies the normalization condition
$\int^\infty_{-\infty}dx\,P(x,t;x_0)=1$ and the initial condition
$P(x,0;x_0)=\delta(x-x_0)$.
The choice of the kernel (\ref{eqn:Kernel}) provides the Langevin
equation (\ref{eqn:ZeroKerneledLangevin})
\begin{equation}
\dot{x}=e^{2S(x)}S'(x)+e^{S(x)}\eta
\label{eqn:NakazatoLangevin}
\end{equation}
with the desired drift force $e^{2S(x)}S'(x)$ which acts as a
restoring force in bottomless regions (that is, for large $|x|$
with $xS'(x)<0$).

One can confirm from Eq.~(\ref{eqn:GreenFunction}) that the
desired distribution $e^{-S}$ is indeed realized, as Tanaka et
al.~expected.
Consider the domain given by
\begin{equation}
D_t=\{x|f^2(x)<2\gamma t\}
\label{eqn:Domain}
\end{equation}
at a given time $t$ with $\gamma\sim1$ a real constant.
Note that $2t$ is the variance of the distribution $P$ as a
function of $f(x)$, and therefore, outside this domain $D_t$, $P$
is considered to be vanishingly small owing to the Gaussian
factor in Eq.~(\ref{eqn:GreenFunction}).
Within $D_t$, on the other hand, this factor is considered to be
of order unity and we have the approximate distribution
\begin{equation}
P(x,t;x_0)\sim\frac{1}{Z_t}e^{-S(x)}
\quad{\mathrm{for}}\quad
x\in D_t,
\label{eqn:ApproximateForm}
\end{equation}
where $Z_t\sim\int_{D_t}dx\,e^{-S(x)}=\int_{D_t}df=\sqrt{8\gamma
t}$.
The desired distribution $e^{-S}$ is seen to be realized in $D_t$
with the appropriate normalization factor $Z_t$.
In other words, the stochastic variable $x$ is considered to be
distributed according to $P\sim e^{-S}/Z_t$, exclusively in $D_t$
at time $t$.

\section{Temporal behavior of the probability distribution}
\setcounter{equation}{0}
In order to confirm the above argument and for illustration, we
consider a typical 0-dim bottomless system given by
$S(x)=m^2x^2/2-\lambda x^4/4$ ($\lambda>0$) and directly study
the behavior of the probability distribution $P$ as a function of
$t$.
Figure~1 displays its temporal behavior, starting from the
initial (ideally delta-shaped) distribution
$P(x,0;x_0)=\delta(x-x_0)$, where $x_0=1.5a$ with
$a=\sqrt{m^2/\lambda}$ being the position of a maximum of the
action.
Note that although the starting point of the stochastic variable
$x=x_0$ is in the bottomless regions (i.e.~$|x|>a$), the
diffusion is well controlled (i.e., $P$ remains in a finite
region even at very large $t$) owing to the restoring force
resulting from the kernel (\ref{eqn:Kernel}).
Furthermore, we can see in Fig.~2, where
$\tilde{S}(x,t;x_0)=-\ln[\sqrt{4\pi t}P(x,t;x_0)]$ [see
Eq.~(\ref{eqn:GreenFunction})] is plotted together with $S$, that
the distribution $P(x,t;x_0)$ is well approximated as $e^{-S}$
within the domain where the stochastic variable is exclusively
distributed.
This is in accordance with the expectation discussed in the
previous paragraph.

Even though the desired Feynman measure $e^{-S}$ seems to be
realized in the course of the stochastic process, the process is
of the diffusion type and has no equilibrium, as is explicitly
seen in the analytical expression (\ref{eqn:GreenFunction}).
The distribution $P$ decays as time increases, and expectation
values over $P$ have no thermal equilibrium limits and may depend
in general on the choice of the initial conditions.
Figure~3 shows the temporal behavior of $\langle
x^2\rangle_t=\int^\infty_{-\infty}dx\,x^2P(x,t;x_0)$ for the same
system as above but with different initial data.
This quantity increases indefinitely and no thermal equilibrium
limit exists.
Note that possible errors that arise in the course of numerical
evaluation of the integrations are very small, typically of order
$10^{-6}$.

It is, however, worth pointing out that the same figure
(Fig.~3) also exhibits a universal behavior of the expectation
values.
After the initial transient time, $\langle x^2\rangle_t$ turns
out to approach a unique curve, albeit monotonically increasing
with $t$, irrespectively of the choice of initial conditions.
This may imply the possibility of extracting ``physics''
underlying bottomless systems through the diffusive stochastic
process.

\noindent
\begin{minipage}{\textwidth}
\begin{center}
\includegraphics[width=0.80\textwidth]{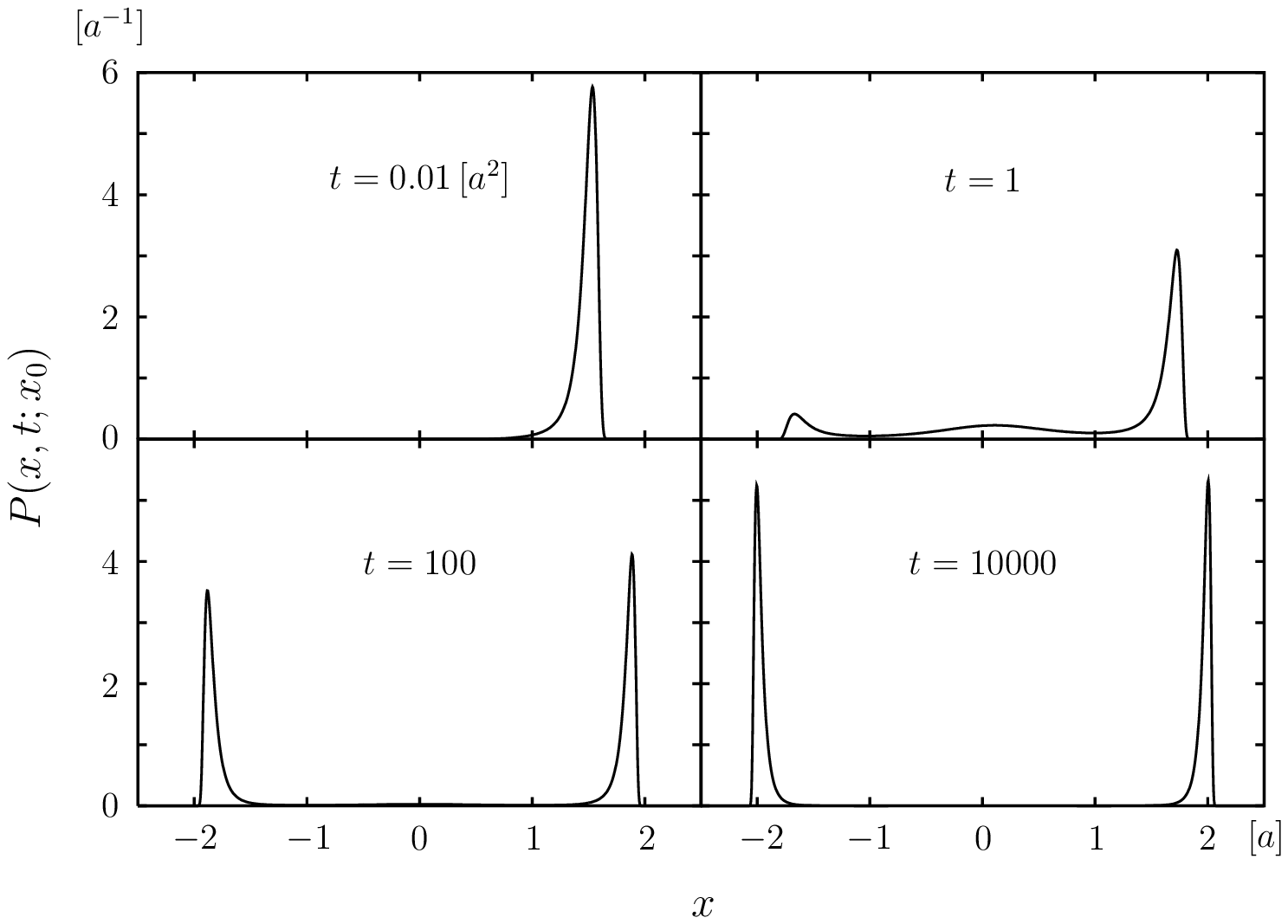}
\end{center}
{\footnotesize Figure 1:
The distribution $P(x,t;x_0)$ for a bottomless system
$S(x)=m^2x^2/2-\lambda x^4/4$ with $S_0=m^4/4\lambda=1.0$ at
$t=0.01,\,1,\,100$ and $10000$ $[a^2]$.
The action has maxima $S_0$ at $x=\pm a=\pm\sqrt{m^2/\lambda}$.
The initial distribution is a delta-shaped function located at
$x_0=1.5a$.}
\end{minipage}
\begin{minipage}{\textwidth}
\vspace*{14mm}
\begin{center}
\includegraphics[width=0.80\textwidth]{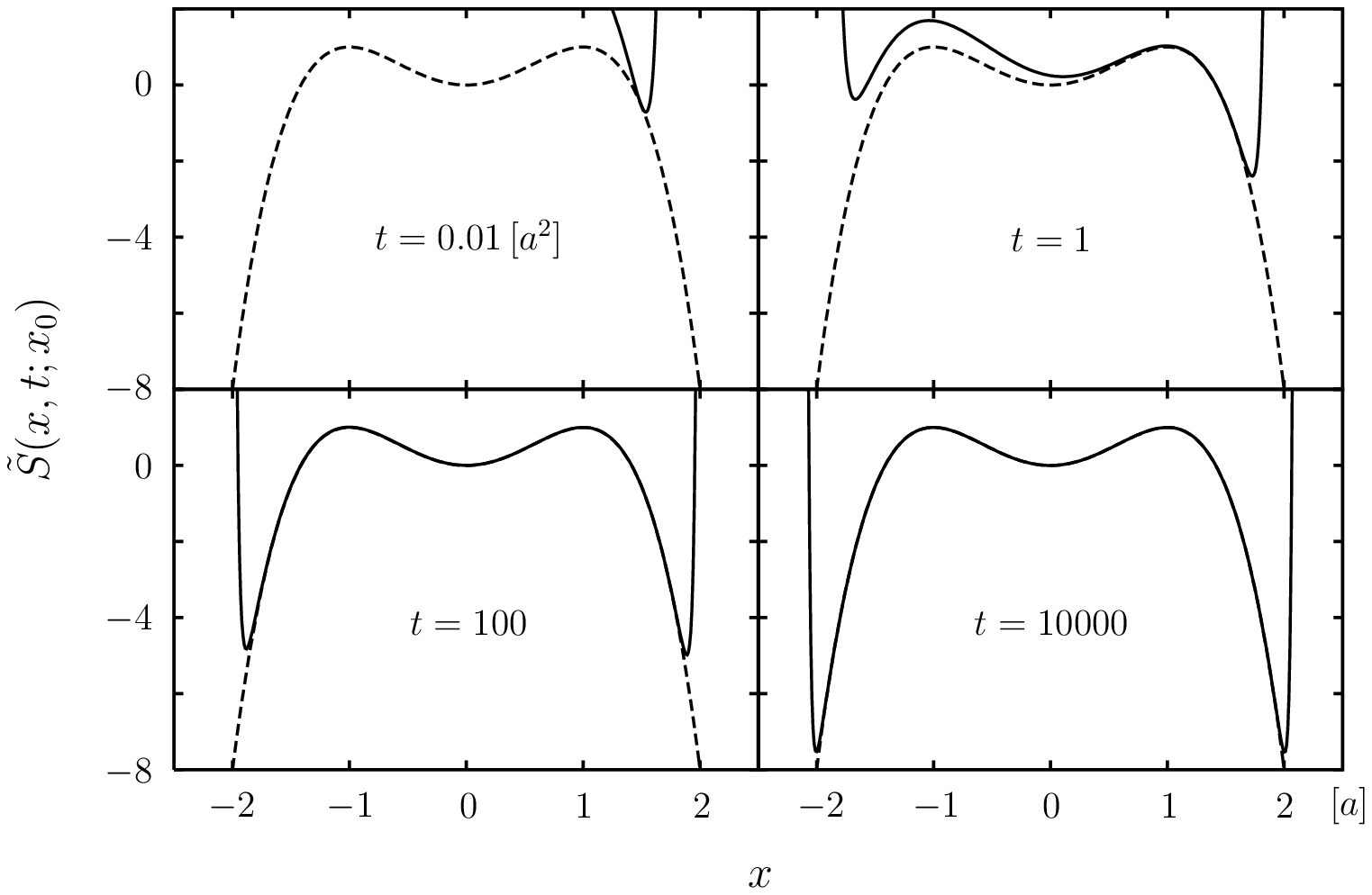}
\end{center}
{\footnotesize Figure 2:
Temporal behavior of $\tilde{S}(x,t;x_0)=-\ln[\sqrt{4\pi
t}P(x,t;x_0)]$ with the initial distribution $\delta(x-1.5a)$.
The dashed curves denote the action $S(x)=m^2x^2/2-\lambda x^4/4$
itself with $S_0=m^4/4\lambda=1.0$.}
\end{minipage}
\begin{minipage}{\textwidth}
\begin{center}
\includegraphics[width=0.80\textwidth]{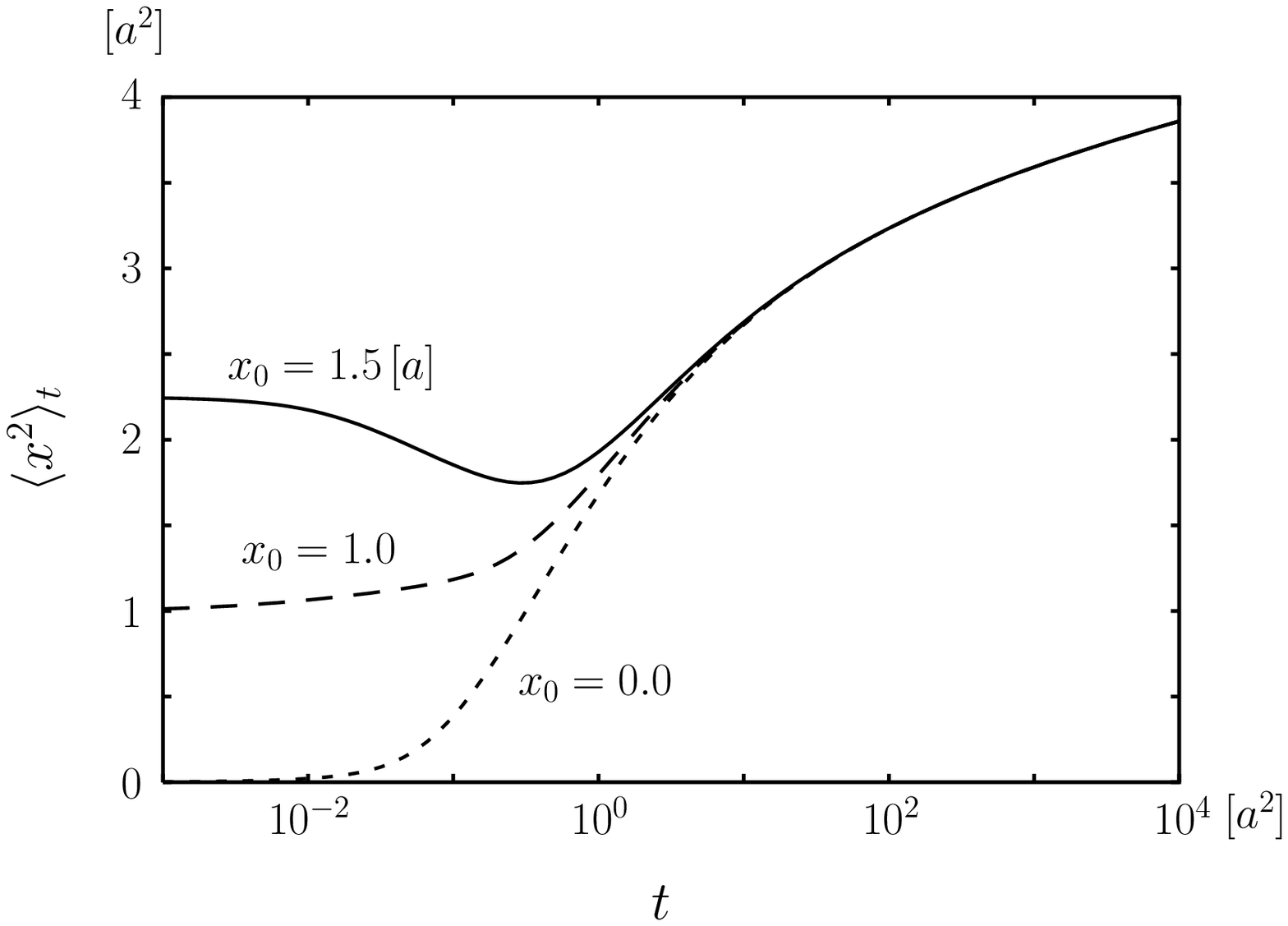}
\end{center}
{\footnotesize Figure~3:
Temporal behavior of the expectation value $\langle x^2\rangle_t$
for the system $S(x)=m^2x^2/2-\lambda x^4/4$
($S_0=m^4/4\lambda=1.0$) with different initial positions
$x_0=0.0,\,1.0$ and $1.5$ $[a]$.}
\vspace*{6mm}
\end{minipage}

\section{Stationary quantities in the diffusive process}
\setcounter{equation}{0}
In order to address this apparently ambitious hope, let us
examine the situation more carefully.
Since we do not know which quantity is to be regarded physically
meaningful for bottomless systems, it seems reasonable for us to
assume that a ``physical quantity'' within the framework of
stochastic quantization is one that has a long-time limit,
irrespective of the initial conditions, for any systems including
bottomless ones.
This is in accordance with the original spirit of stochastic
quantization and may be considered a minimal requirement.
[Of course, this is just one possibility, and our assumption is
not meant to exclude other possibilities, such as that presented
in Ref.~\cite{ref:Greensite}.]
For diffusive stochastic processes, however, whether such a
quantity exists or not is a highly nontrivial question, because
we naturally anticipate that every quantity should have a trivial
(either vanishing or diverging) large-time limit.
The idea here is to discriminate different dynamics possibly
coexisting in such diffusion processes.
In other words, our task is to extract appropriate ``physics''
underneath the dominating diffusive process caused by the
unboundedness of the classical action.
The following analysis is a first step along this line of
thought.

Observe that the domain $D_t$ defined by Eq.~(\ref{eqn:Domain})
expands monotonically as $t$ increases.
This represents one of the main features of diffusion.
We attempt to estimate this expansion quantitatively.
Let $x_+(t)$ [$x_-(t)$] be the right(left) end of the domain
$D_t$:
\begin{equation}
\int^{x_\pm(t)}_{x_0}dy\,e^{-S(y)}=\pm\sqrt{2\gamma t}.
\label{eqn:Boundary}
\end{equation}
[See Eqs.~(\ref{eqn:Domain}) and (\ref{eqn:GreenFunction}).]
If we change the initial position $x_0$ by an infinitesimal
amount $\delta x_0$, it yields the following changes in
$x_\pm(t)$:
\begin{equation}
\delta x_\pm=e^{S(x_\pm)-S(x_0)}\delta x_0.
\end{equation}
This shows that $\delta x_\pm$ become vanishingly small as time
increases, since $x_+$ increases and $x_-$ decreases
monotonically in time and therefore $e^{S(x_\pm)}\to0$.
The difference between the domains $D_t$ for different values of
$x_0$ then disappears.
This implies an approach to a unique distribution $P\propto
e^{-S}$.
This is the reason that we observe a unique curve for expectation
values at large times in Fig.~3.
It is important to note that the expansion of $D_t$ is well
controlled in this stochastic process.
In fact, we can easily see, from Eq.~(\ref{eqn:Boundary}),
\begin{equation}
\dot{x}_\pm
=\pm\sqrt{\frac{\gamma}{2t}}e^{S(x_\pm)}
\sim\pm\frac{2\gamma}{Z_t}e^{S(x_\pm)},
\label{eqn:xpmdot}
\end{equation}
which implies that the expansion rate diminishes exponentially at
large $t$ (or large $|x_\pm|$).

This observation may enable us to extract a quantity that has a
long-time limit.
Let us examine the time development of the expectation value of
an arbitrary function $F(x)$, $\langle
F(x)\rangle_t=\int^\infty_{-\infty}dx\,F(x)P(x,t;x_0)$.
Using the Fokker--Planck equation
(\ref{eqn:ZeroKerneledFokkerPlanck}), we obtain its time
derivative:
\begin{eqnarray}
&&\frac{d}{dt}\langle F(x)\rangle_t
=\int^\infty_{-\infty}dx\,F(x)
\frac{\partial}{\partial x}e^{2S(x)}\left(
\frac{\partial}{\partial x}+S'(x)
\right)P(x,t;x_0)\nonumber\\
&&\nonumber\\
&&\phantom{\frac{d}{dt}\langle F(x)\rangle_t}
=\int^\infty_{-\infty}dx\,\left(
\frac{\partial}{\partial x}e^{S(x)}F'(x)
\right)e^{S(x)}P(x,t;x_0).
\end{eqnarray}
This can be rewritten, under the validity of the approximation
(\ref{eqn:ApproximateForm}), as
\begin{equation}
\frac{d}{dt}\langle F(x)\rangle_t
\sim\frac{1}{Z_t}\int_{D_t}dx\,
\frac{\partial}{\partial x}e^{S(x)}F'(x)
=\left.\frac{1}{Z_t}e^{S(x)}F'(x)\right|^{x_+(t)}_{x_-(t)}.
\label{eqn:EvolutionF}
\end{equation}
With the help of the equations for $x_+(t)$ and $x_-(t)$ in
Eq.~(\ref{eqn:xpmdot}), it can be further reduced to
\begin{equation}
\frac{d}{dt}\langle F(x)\rangle_t
\sim\frac{1}{2\gamma}\Bigl(
\dot{x}_+F'(x_+)+\dot{x}_-F'(x_-)
\Bigr)
=\frac{1}{2\gamma}\left(
\frac{d}{dt}F(x_+)+\frac{d}{dt}F(x_-)
\right).
\end{equation}
One may thus expect that the quantity $\langle\!\langle
F(x)\rangle\!\rangle_t$, defined by
\begin{equation}
\langle\!\langle F(x)\rangle\!\rangle_t
=\frac{1}{2\gamma}[F(x_+(t))+F(x_-(t))]-\langle F(x)\rangle_t,
\label{eqn:<<F(x)>>}
\end{equation}
\begin{minipage}{\textwidth}
\begin{center}
\includegraphics[width=0.80\textwidth]{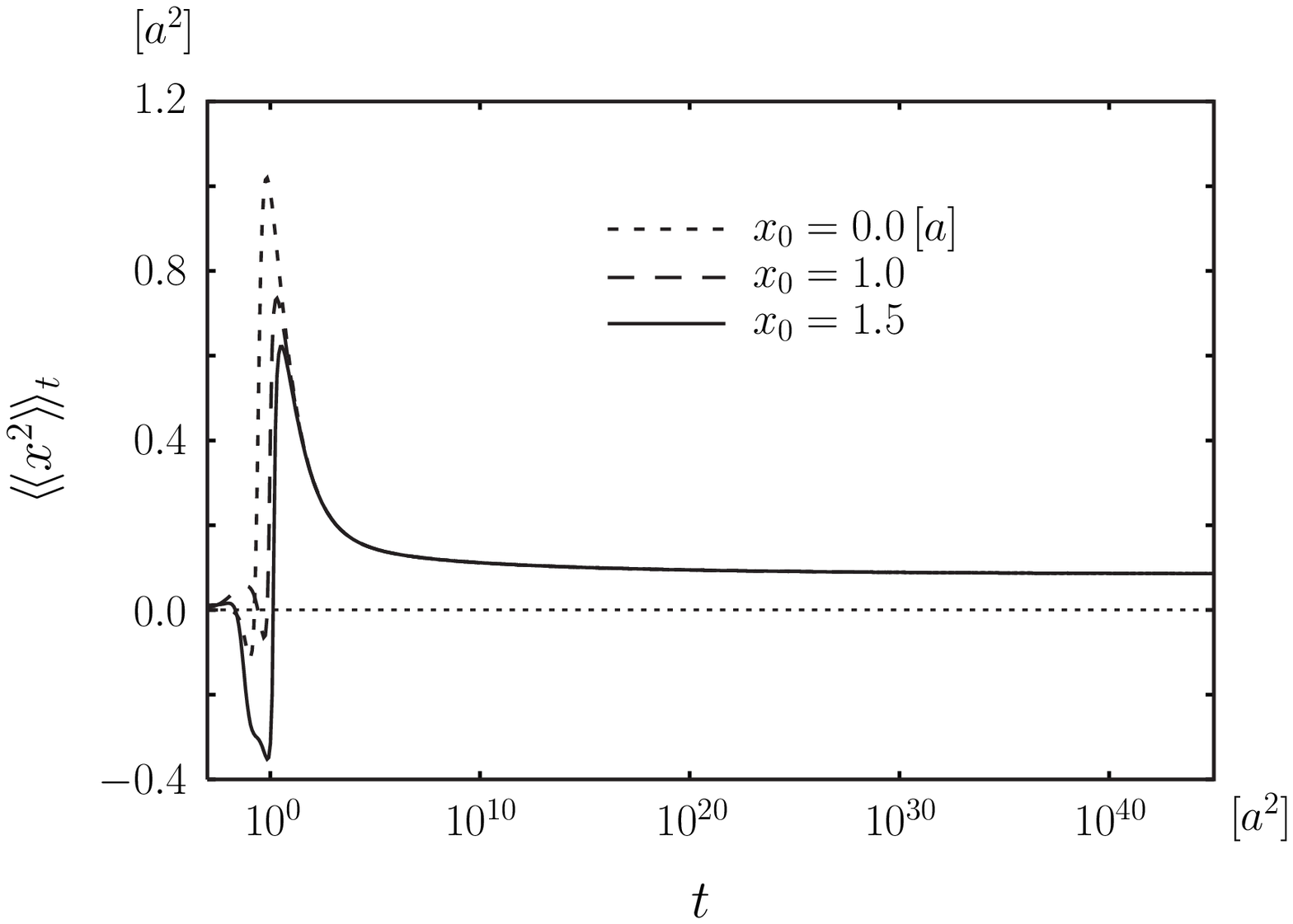}
\end{center}
{\footnotesize Figure~4:
Temporal behavior of $\langle\!\langle x^2\rangle\!\rangle_t$ for
a bottomless system $S(x)=m^2x^2/2-\lambda x^4/4$ with
$S_0=m^4/4\lambda=1.0$.
Initial positions are chosen as $x_0=0.0,\,1.0$ and $1.5$ $[a]$.
The parameter $\gamma$ is chosen to be $0.995$.}
\vspace*{12mm}
\end{minipage}
has a $t$-independent value for large $t$.
Roughly speaking, $\langle\!\langle F(x)\rangle\!\rangle_t$ is a
quantity reflecting the ``quantum fluctuations'' in the
bottomless system, since the first two terms in the square
parentheses represent ``deterministic'' contributions.
[Remember that the dynamics for $x_\pm(t)$ in
Eq.~(\ref{eqn:xpmdot}) are deterministic.]

The above argument has been confirmed by the numerical
calculation of $\langle\!\langle x^2\rangle\!\rangle_t$ for the
system $S(x)=m^2x^2/2-\lambda x^4/4$ whose results are given in
Fig.~4.
(Here errors are typically of order $10^{-4}$.)
These results show that after the initial transient time, the
quantity $\langle\!\langle x^2\rangle\!\rangle_t$ approaches a
certain value, which is constant over a wide range of large $t$,
irrespectively of the choice of the initial position $x_0$.
Similar behavior is observed for other correlation functions.

\section{Discussion}
\setcounter{equation}{0}
The observation of a time-independent quantity in a diffusive (or
diverging) stochastic process is surprising and worthy of note.
We hope that it will lead to further insight for bottomless
systems.
The above quantity $\langle\!\langle
F(x)\rangle\!\rangle_{t\to\infty}$ is a candidate for the
time-independent quantity, representing a kind of quantum effect
in such systems.
It can be thought, as described above, to be a measure of the
quantum fluctuations for bottomless systems.

We must remember, however, that the present analysis is largely
dependent on our choice of the specific kernel function $K(x)$
(\ref{eqn:Kernel}).
In particular, the asymptotic value of $\langle\!\langle
F(x)\rangle\!\rangle_t$ possibly depends on the very choice of
the kernel.
Even though it seems plausible that for bottomless systems
meaningful quantities are rather limited, and accordingly not all
conceivable kernels can provide us with them, it is very
important to clarify this point, e.g.~by determining the kernel
dependence of $\langle\!\langle
F(x)\rangle\!\rangle_{t\to\infty}$.
This is an important issue to be studied, though a different
choice of the kernel no longer ensures the solvability of the
Fokker--Planck equation and may entail an additional numerical
study in order to obtain the probability distribution $P$.
Other points worth further investigation include
(a) the relevance (if any) of $\langle\!\langle
F(x)\rangle\!\rangle_t$ to observable quantities,
(b) formulation on the basis of the corresponding Langevin
equation (\ref{eqn:NakazatoLangevin}),
(c) comparison with other proposals \cite{ref:Greensite}, and
(d) extension to higher dimensional systems.
Work is in progress on these points.

\section*{Acknowledgements}
The authors acknowledge useful and helpful discussions with
Professor I.~Ohba and Dr.~Y.~Yamanaka.
They also thank Helmuth H\"uffel for critical reading of the
manuscript and discussions and Professor M. Namiki for
enlightening suggestions.
This work is supported by a Grant-in-Aid for JSPS Research
Fellows.


\end{document}